\documentclass[%
 prl,
superscriptaddress,
twocolumn,
 amsmath,amssymb,
 aps,
]{revtex4-2}

\usepackage{graphicx}
\usepackage{dcolumn}
\usepackage{bm}

\usepackage[colorlinks,linkcolor={blue},citecolor={blue},urlcolor={blue}]{hyperref}

\usepackage{eqnarray,amsmath}
\newcommand{\nhat}{\hat{\bf n}}

\newcommand{\xhat}{\hat{\bf x}}
\newcommand{\yhat}{\hat{\bf y}}

\newcommand{\rb}{{\bf r}}
\newcommand{\ub}{{\bf u}}

\newcommand{\Fb}{{\bf F}}
\newcommand{\fb}{{\bf f}}

\newcommand{\Ib}{{\bf I}}

\newcommand{\Tb}{{\bf T}}

\newcommand{\vb}{{\bf v}}

\newcommand{\Omegab}{\mbox{\boldmath $\Omega$\unboldmath}}
\newcommand{\omegab}{\mbox{\boldmath $\omega$\unboldmath}}

\newcommand{\tauhat}{\hat{\mbox{\boldmath $\tau$\unboldmath}}}

\newcommand{\ellhat}{\hat{\mbox{\boldmath $\ell$\unboldmath}}}

\newcommand{\nablab}{\mbox{\boldmath $\nabla$\unboldmath}}

\newcommand{\beq}{\begin{equation}}

\newcommand{\eeq}{\end{equation}}

\newcommand{\bea}{\begin{eqnarray}}

\newcommand{\eea}{\end{eqnarray}}

\usepackage[colorinlistoftodos]{todonotes}
\newcommand{\rhob}{\mbox{\boldmath $\rho$\unboldmath}}

\usepackage{orcidlink}

\begin{document}


\title{Shape-asymmetry and flexibility in active cross-stream migration in nonuniform shear}

\author{Derek C. Gomes\orcidlink{0009-0006-6318-2624}}
\email{derekgomes@students.iisertirupati.ac.in}
\author{Tapan C. Adhyapak\orcidlink{0000-0002-8251-2880}}%
\email{Author to whom correspondence should be addressed: adhyapak@labs.iisertirupati.ac.in}
\affiliation{Department of Physics, Indian Institute of Science Education and
Research (IISER) Tirupati, Tirupati, Andhra Pradesh, 517619, India }


\begin{abstract}

We show that activity and broken fore-aft shape symmetry enable
microswimmers to cross streamlines in nonuniform shear, a key yet overlooked
factor in active cross-stream migration. Using a model of flagellated
microswimmers in microchannel flow, we find that hydrodynamic coupling and
flagellar flexibility significantly impact migration. A simplified theory
identifies key factors driving the underlying rich nonlinear dynamics.  Our
findings apply to dynamics and control of both living and artificial
microswimmers, while the hydrodynamic framework extends to diverse shear flow
scenarios.

\end{abstract}

\maketitle


Most theoretical studies on active suspensions assume suspended active
particles to be rigid and head-tail symmetric (HTS) in shape
\cite{Saintillan_active_rheo, Speck_activeBrownian, Lowen_dense_coll,
Menzel_swimmer_in_LiqCryst, marchetti_RMP, Lusi_Goldstein_self_organise,
Lopez_superfluid}. This simplifies their dynamics by decoupling them
from self-induced active flows, while coupling to external flows remains
centrosymmetric and non-extended \cite{nonextended}.

\begin{figure}

\includegraphics[height=0.25\columnwidth]{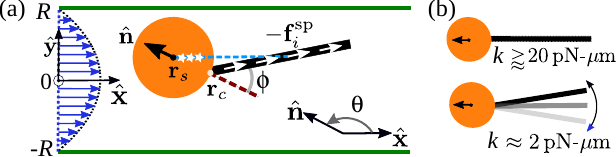}

\caption{(a) A model microswimmer with broken head-tail and axial symmetries in
Poiseuille flow. Stars represent distributed hydrodynamic image singularities
corresponding to one of the active forces, $-\fb^{\rm{sp}}_i$.  (b) Schematic
illustrating the `rigid' and `flexible' limits of the flagella in terms of the
bending rigidity modulus $k$.}

\label{Fig1}

\end{figure}

This paper examines the dynamics of a model microswimmer with broken
head-tail and axial shape-symmetries, considering its extended coupling to an
imposed Poiseuille flow. The model features a finite-sized cell body and a
flexible flagellar bundle, represented by a slender rod bending at the
flagellar joint [see Fig. \ref{Fig1}].  Finely resolved hydrodynamics capture
the active-flow mediated coupling between the cell body and flagella, while
external shear flow affects them unevenly. The study reveals dynamics distinct
from rigid HTS microswimmers in Poiseuille flows and suggests novel control
strategies for flagellated microswimmers in microchannels with potential
biomedical and biotechnological applications.

An HTS microswimmer in Poiseuille flow oscillates around streamlines while
maintaining average streamline motion \cite{Benyahu_larva_in_flow,
zottl_prl_NLD_channel, salima_prl, zottl2013periodic,
Chacon_Chaotic_Poiseuille}. In contrast, here we show that head-tail asymmetric
(HTA) swimmers mostly migrate to channel walls, except for those in a small
phase-space region bounded by an \emph{unstable limit cycle}, which move toward
the channel \emph{midsection}. While the \emph{extended} hydrodynamic coupling to the
imposed flow is crucial, flagellar flexibility further aids in avoiding
the midsection, preventing flushing out by the channel flow.

Most suspended active particles exhibit pronounced HTA shapes. Notable
examples include flagellated microswimmers like \emph{E. coli} and \emph{B.
subtilis}, which are increasingly used as prototypical \emph{living} active
particles \cite{Weak_sync_Wu, Chate_nematic_bac, Poon_ecoli_model,
Bar_bac_phase_diag, Tailleur_multi_bac, Polin_bac_phase_trans,
Wu_active_interface, Ariel_active_entropy}. Advances in engineered active
matter also enable artificial microswimmers with flexible appendages
\cite{Dreyfus_artificial_swimmer, Golestanian_act_coll, Paxton_act_nanorods,
Palacci_light_act, Golestanian_magnetic_act, Ambarish_magnetic_active,
Thutupalli_drop_squirm}. The flagellar bundles forming their tails
significantly modulate cell body dynamics and cannot always be ignored
\cite{Ardekani_flag_buckling, Aranson_flagella_bending, Aranson_mucus,
Mathijssen_rheo_nonnewton, Lauga_flag_flow, tapanSM2016}. Here, we analyze the
fundamental impact of flagellar motion on overall cell dynamics -- an aspect
rich in complexity yet insufficiently understood.

Microswimmer cross-streaming, previously predicted and demonstrated, has
been explained using HTS models, attributing it to activity and
shape-independent factors such as inertial lifts \cite{Ardekani_migration,
paper7_akash_nature}, medium viscoelasticity \cite{Luca_visco_migration,
akash_SM2022}, wall interactions \cite{paper18_akash_epj,
zottl_prl_NLD_channel, Hill_upstream}, field alignment \cite{Kessler_gyrotaxis,
salima_prl, ranabir_PRL2024}, and unsteady flagellar beating
\cite{omori_rheotaxis}. In contrast, we analyze the role of \emph{extended}
hydrodynamic coupling in active HTA models, demonstrating its dominance in
experimentally relevant conditions. The predicted cross-streaming for active
HTA particles is significantly stronger than that observed for passive
bacterial-sized particles \cite{paper24_aranson_rsif} and stems from a
fundamentally different mechanism than those governing other passive
cross-streaming systems \cite{Bretherton_jeff_orbit, *segre_silberberg,
*Zimmerman_Passive_cross_stream, *holger_polymer_2011}.

Our model (Fig. \ref{Fig1}) is based on a flagellated bacterium with a
rigid cell body and a flexible flagellar bundle. To capture finite size and HTA
shape minimally, we model the cell body as a sphere of radius $a$ centered at
$\rb_{\rm{S}}$ and approximate the flagellar bundle as a slender rod of length
$\ell_f$ oriented along $\ellhat_f$. One end of the rod attaches to the cell
body at $\rb_c = \rb_{\rm{S}} - a \nhat$, where $\nhat$ is the cell's intrinsic
orientation. Flagellar bending from $\nhat$
\cite{turner_bend} is captured minimally by allowing a bend angle $\phi$, which
is opposed by an \emph{elastic} restoring torque $\Tb^{\rm{el}}_f=-k\phi
\tauhat$, where $k$ is the bending rigidity modulus and $\tauhat =
\ellhat_f\times\nhat/\|\ellhat_f\times\nhat\|$ is a unit vector normal to both
$\nhat$ and $\ellhat_f$.

As in flagellated bacteria, we assume the rod pushes back on the ambient
fluid by applying point forces $\{-\fb^{\rm{sp}}_i\}, i=1$ to $N$, at $N$
locations along the rod.  The reactionary self-propulsion force $\Fb^{\rm{sp}}
= \sum_i \fb^{\rm{sp}}_i$ is transferred to the cell body at $\rb_c$. The model
operates at a constant active force strength $F^{\rm{sp}}$, ensuring constant
activity \cite{marchetti_RMP}. However, due to flagellar bending, it
dynamically adjusts its translational and rotational self-propulsion velocities
in response to environmental factors ({\it e.g.}, crowds or imposed flows).

\textit{Equations of motion -- } At low Reynolds numbers \cite{Dhont}, the
swimmer's overdamped dynamics are governed by the force and torque balance
conditions:
\begin{eqnarray}
&\Fb_s + \Fb_f + \Fb^{\rm{sp}} = 0, \label{force_bal}\\
&\Tb_s + \Tb_f + \Tb^{\rm{sp}} = 0, \label{torque_bal}
\end{eqnarray}
which account for all forces and torques from the fluid, the latter evaluated
at $\rb_c$ for simplicity.  Here, $\Fb_s = -6\pi\eta a(\vb_s -\vb_s^0)$ and
$\Tb_s = -8\pi\eta a^3(\omegab_s -\omegab_s^0) + a(\nhat\times \Fb_s)$ are,
respectively, the viscous force and torque on the cell body, while the
corresponding contributions on the flagella, predicted by slender rod theory
\cite{pozrikidis, paper32_lauga_JFM}, are
\begin{eqnarray}
\Fb_f &=& -\int_0^{\ell_f} \rhob\cdot \left[\vb_f(s) -\vb_f^0(s)\right]\, ds, \\
\Tb_f &=& -\int_0^{\ell_f} s\ellhat_f \times \rhob \cdot\left[\vb_f(s) -\vb_f^0(s)\right]\, ds. \label{eq:torq}
\end{eqnarray}
Here, $\eta$ is the fluid viscosity, and $\rhob =
\rho_{\parallel}\ellhat_f\ellhat_f + \rho_{\perp}(\Ib - \ellhat_f\ellhat_f)$ is
the resistivity tensor \cite{Chwang_Wu_1975}; $\vb_s$ and $\omegab_s$ are
respectively the translational and rotational velocities of the cell body,
while $\vb_f(s) = \vb_s - a(\omegab_s\times\nhat) + s\left(\dot{\phi}\tauhat\times
\ellhat_f\right)$ is the flagellar velocity at a distance $s$ from $\rb_c$;
$\{\vb_s^0, \omegab_s^0, \vb_f^0(s)\}$ denote the flow-induced velocities (see
below), while $\Fb^{\rm{sp}}$ and $\left.  \Tb^{\rm{sp}}\right|_{\rb_c}(=0)$
represent the active force and torque on the swimmer, respectively.
Furthermore, the internal elastic torque $\Tb^{\rm{el}}_f$ and its reaction
$\Tb^{\rm{el}}_s= -\Tb^{\rm{el}}_f$ on the cell body, enter the dynamics
through the torque balances on the sphere and the rod, taken separately: $\Tb_s
= - \Tb^{\rm{el}}_s = - k\phi\tauhat$ \& $\Tb_f =
- \Tb^{\rm{el}}_f= k\phi\tauhat$.

\begin{figure*}

\includegraphics[height=0.35\textwidth]{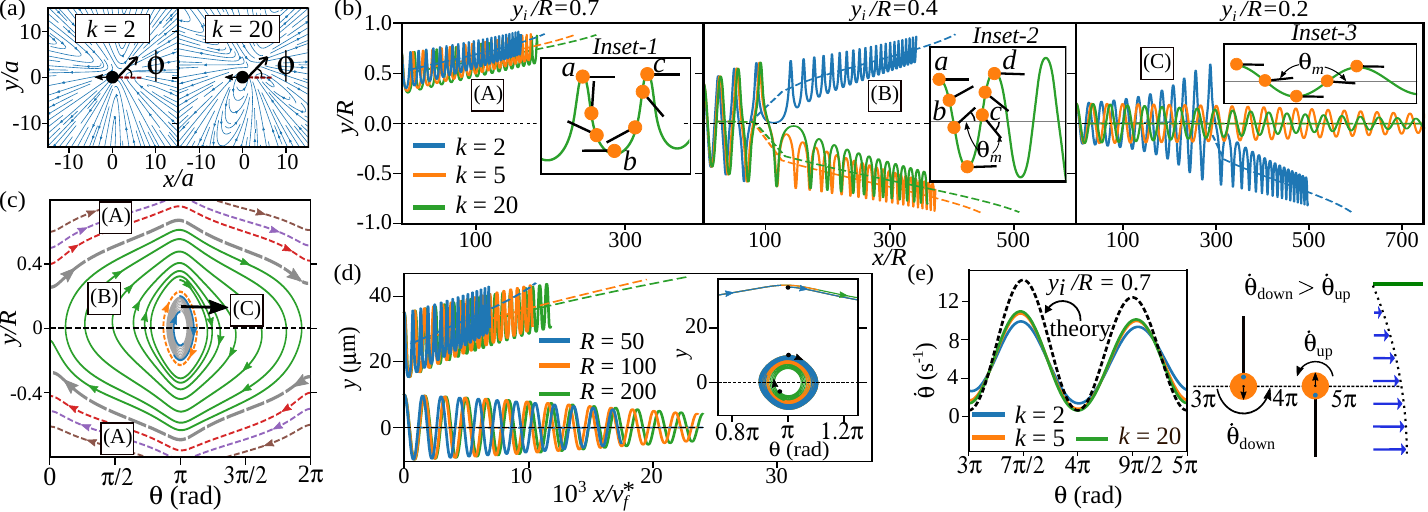}

\caption{(a) The intrinsic flow field of our model microswimmer for
$\phi=45^{\circ}$ in the flexible ($k=2$) and rigid ($k=20$) limits.  (b) Three
distinct trajectory classes [Class (A), (B), and (C)] in a Poiseuille flow for
varying $k$, with schematic swimmer orientations in the insets. The flow and
trajectories run left to right. $R = 50\, \mu\rm{m}$, $v_f = 500\,
\mu\rm{m}/\rm{s}$.  (c) Phase space trajectories of a rigid ($k=20$)
\emph{pusher} in the $y$-$\theta$ plane, showing an \emph{unstable limit cycle}
(orange curve). Flexible swimmer and puller cases are discussed in the text.
(d) Similarity of trajectories for varying $R$ and $v_f$; $v_f/R^2=0.2$,
$k=20$.  \emph{Inset:} Overlapping phase space trajectories.  (e)
$\dot{\theta}$ vs.  $\theta$ plot of Class-(A) trajectories, from theory for
rigid swimmers at $y/R=0.7$ (dashed line) and from simulations for varying $k$
at $y_i/R=0.7$ (continuous lines); the range $\theta \in [3\pi,5\pi]$ is chosen
to cover one period while avoiding end effects.  \emph{Schematic} illustrates
nonuniform $\dot\theta$ for \emph{up} vs. \emph{down} orientations, with blue
dots marking the \emph{shifted} hydrodynamic center.  }

\label{Fig2}
\end{figure*}

\textit{Finely resolved hydrodynamics --} To evaluate the
\emph{flow-induced} velocities, we consider the flow field around the swimmer:
$ \ub(\rb) = \ub_f^{\rm{active}}(\rb) + \ub_f^{\rm{mov}}(\rb) +
\ub_s^{\rm{mov}} (\rb)+ \ub^{\rm{shear}}(\rb) +
{\rm{Img}}[\ub_f^{\rm{active}}(\rb) + \ub_f^{\rm{mov}}(\rb)+
\ub^{\rm{shear}}(\rb)]$.  This field comprises contributions from the flagellar
active forces ($\ub_f^{\rm{active}}$), flagellar movement ($\ub_f^{\rm{mov}}$),
cell body's motion ($\ub_s^{\rm{mov}}$), and imposed shear
($\ub^{\rm{shear}}$). Clearly, from \eqref{force_bal}-\eqref{eq:torq}, the
fields $\ub_f^{\rm{mov}}, \ub_s^{\rm{mov}}$ implicitly contain contributions of
active forces on both cell body and flagella.  The image fields (`Img') account
for the flows \emph{reflected} off the cell body, satisfying the stick boundary
condition.  These contributions are  detailed in the Supplemental Material
\cite{supplemental_material}.  The flow-induced velocities, mentioned above,
are then given by:  $ \vb_s^0 = \left[1 + (a^2/6)\nabla^2\right]\ub_s^0(\rb_s),
\omegab_s^0 = (1/2)\nablab \times \ub_s^0(\rb_s), $  where $\ub_s^0 =
\ub_f^{\rm{mov}} + \ub^{\rm{active}} + \ub^{\rm{shear}}$ is the flow field in
the absence of the sphere, and $\vb_f^0(s) = \ub(s)$ along the slender rod.

Thus, the velocities $\{\vb_s^0, \omegab_s^0, \vb_f^0(s)\}$ depend on the
flow-fields, which in turn depend on the cell body and flagellar movements.
Using their expressions, \eqref{force_bal}-\eqref{torque_bal} can be solved
self-consistently to obtain $\vb_s, \omegab_s$ and
$\Omegab_f=\dot{\phi}\tauhat$ \cite{supplemental_material}. The dynamics of the
swimmer are then described by $d\rb_s/dt = \vb_s, d\nhat/dt = \omegab_s\times
\nhat$ and $d\ellhat_f/dt = \Omegab_f\times \ellhat_f$.

In the following, we take $a = 1\, \mu\text{m}$, $\ell_f = 5a$, and
$(\eta, \rho_\parallel, \rho_\perp) = (0.89, 3.9, 6.7) \times 10^{-3}\,
\text{Pa}\text{-}\text{s}$, based on flagellated bacteria in room-temperature
water \cite{tapanPRE2015}. We adjust $F^{\rm{sp}}$ so that $v_s=50\,
\mu\rm{m}/\rm{s}$ \cite{rusconi2014bacterial}, and consider $k = 2, 5,$ and
$20\, \text{pN}\text{-}\mu\text{m}$ unless otherwise stated, representing
flexible, intermediate, and rigid flagellar bundles, respectively
\cite{supplemental_material}. We omit mentioning the unit of $k$ henceforth for
brevity.

Figure \ref{Fig2}(a) displays the flow field $\ub(\rb)$ in the absence of
shear for a bending angle $\phi = 45^{\circ}$, comparing two flagellar
rigidities: $k = 2$ (flexible) and $k = 20$ (rigid). In both cases, $\ub(\rb)$
is asymptotically a \emph{stresslet} field (off-aligned to $\nhat$), but the
proximal fields are significantly altered by non-axisymmetry.

\textit{Dynamics in non-uniform shear --}  We analyze the swimmer's
dynamics in a Poiseuille flow, given by $\ub^{\rm{shear}} = v_f(1 -
y^2/R^2)\xhat$, confined between two parallel walls at $y = \pm R$ [Fig.
\ref{Fig1}(a)].  Considering high Péclet number dynamics across the channel
width, we focus on the deterministic motion of the swimmer in the $xy$ plane.
Since shear-induced alignment is predominantly along the \emph{shear plane}
\cite{shear-induced}, we assume an in-plane orientation for simplicity. The
channel parameters are set to $R = 50\, \mu\text{m}$ and $v_f = 500\,
\mu\text{m}/\text{s}$ \cite{rusconi2014bacterial, junot2019swimming}, though
our results remain valid for other values as well [see Fig. \ref{Fig2}(d) and
Appendix C].

Our main observations are presented in Figs. \ref{Fig2}(b)-(d). We first
discuss the results for a pusher with rigid flagella ($k = 20$). Fig.
\ref{Fig2}(b) shows the trajectories traced by $\rb_s$, starting from an
initial upstream orientation ($\theta_i = \pi$) for various initial vertical
positions $y_i$, of the swimmer.  All trajectories are downstream and
oscillatory. Remarkably, the mean paths (colored dashed lines) exhibit a
striking cross-stream migration, contrasting the net \emph{streamline} motion
observed in HTS swimmers \cite{zottl_prl_NLD_channel, salima_prl,
zottl2013periodic}.  The trajectories can be categorized into three classes
associated with distinct ranges of $y_i$: (A) Trajectories with large $y_i$
oscillate away from the channel midsection (black dashed line),
gradually approaching the nearest wall (\textit{Panel-1}).  (B) Trajectories
with intermediate $y_i$, above a threshold ($y_{\rm{max}}$), initially
oscillate about the midsection with \emph{growing} amplitudes
(\textit{Panel-2}). The swimmer is thus temporarily `trapped' to the midsection
on average until it completes a full period within one half of the channel,
after which it `escapes' toward the wall on that half.  (C) Trajectories with
$y_i < y_{\rm{max}}$ oscillate about the midsection with \emph{decaying}
amplitudes, causing the swimmer to remain trapped along the midsection
indefinitely (\textit{Panel-3}).

The corresponding $\theta$-dynamics in the above classes are also distinct. In
(A) $\nhat$ rotates continuously, following the fluid vorticity
(counterclockwise above and clockwise below the midsection, respectively).
Thus, the cell tumbles, completing one full rotation approximately within one
period of the spatial trajectory (\emph{Inset-1}). In contrast, in (B) \& (C),
$\nhat$ oscillates between $\theta=\pi \pm \theta_m$, with $\theta=\pi$ marking
the approximate turning points (crests and troughs of the spatial
oscillations), while $\theta=\pi \pm \theta_m$ is reached near the midsection
(\emph{Inset-2}). As in the spatial oscillations, the $\theta$-oscillations in
(B) have growing amplitudes, $\theta_m(t)$, whereas those in (C) decay
over time. Accordingly, $\nhat$ in (B) starts tumbling once the trajectory
`escapes' the midsection. In contrast, in (C), $\nhat$ tends toward a
steady-state upstream orientation, $\theta=\pi$, while the swimmer remains
trapped to the midsection. This indicates that $(y, \theta)=(0,\pi)$ is one of
the system's \emph{dynamical attractors}.

The effect of the full range of $\theta_i$ can be read off from the phase
space diagram in the $y$-$\theta$ plane, plotted in Fig.  \ref{Fig2}(c). The
diagram reveals an \emph{unstable limit cycle} (dashed orange curve) that
separates the growing- and decaying-amplitude trajectory classes, (B) and (C),
respectively.  Additionally, a separatrix (dashed gray curve) partitions
classes (A) and (B).

Our results are universal across varying channel parameters, as shown in
Fig. \ref{Fig2}(d). Trajectories in channels with different $R$ are similar
when $v_f/R^2$ is kept constant and $x$ is scaled by the nondimensionalized
maximum flow speed, $v_f^* = v_f/(F^{\rm{sp}}/6\pi\eta a)$. Notably, the
corresponding curves perfectly overlap in the $y$-$\theta$ plane, indicating an
invariant limit-cycle size [inset, Fig. \ref{Fig2}(d)].

\textit{Anomalous channel dynamics and HTA --}  The oscillatory
trajectories of an HTS swimmer in Poiseuille flow are attributed to the
tumbling of its self-propulsion direction due to flow-induced rotations
\cite{zottl_prl_NLD_channel, salima_prl, zottl2013periodic}.  Although HTA
swimmer oscillations share the same origin, the contrasting net vertical
migration (along $\pm \yhat$) observed here arises from the asymmetric
influence of the flow along the swimmer's body.

For a perfectly rigid swimmer, $\ellhat_f = -\nhat$. Considering only the
dominant flow contribution, $\ub^{\rm{shear}}$, along the flagellar
background, the equations for $\dot{\rb}_s = (\dot{x}, \dot{y})$ and
$\dot{\nhat} = (-\sin{\theta}, \cos{\theta}) \dot{\theta}$ yield:  
\begin{eqnarray}
&&\dot{y} = v^{\rm{sp}} \sin{\theta} + r_{\rm{sh}} \dot{\theta} \cos{\theta} + \frac{v_f F(y,\theta)}{2R^2} \sin{2\theta}, \label{eq:ydot}\\
&&\dot{\theta} = \frac{v_f}{R^2} \left[\alpha y + 2\beta_2 \sin^2{\theta} \,y + \beta_1 \sin{\theta} + \beta_3\sin^3{\theta}\right]. \label{eq:thetadot}
\end{eqnarray}  
Here, $v^{\rm{sp}} = F^{\rm{sp}} / (6 \pi \eta a + \rho_{\parallel} \ell_f)$,
and the other coefficients are listed in \cite{supplemental_material}. The
coefficient $r_{\rm{sh}} = \rho_{\perp} \ell_f (a + \ell_f/2) / (6 \pi \eta a +
\rho_{\perp} \ell_f)$ represents the shift of $\rb_s$ from the \emph{hydrodynamic center}.  Note that for an HTS swimmer, $r_{\rm{sh}}, F(y,\theta), \beta_1,
\beta_3 = 0$, $\alpha=1$, which we recover here in the limit
$\ell_f \to 0$ \cite{supplemental_material},  while $\beta_2$ is the familiar Jeffery
coefficient \cite{zottl2013periodic}.  Further, the shape-HT symmetry is
purely broken by the $\beta_1$ and $\beta_3$ terms alone.

Eqs. \eqref{eq:ydot} and \eqref{eq:thetadot} describe a $y$-$\theta$
dynamics that is decoupled from $x$ and dependent on $v_f$ and $R$ only through
the ratio $v_f/R^2,$ the flow angular velocity gradient. This explains the
\emph{similarity} factor used in Fig.  \ref{Fig2}(d).  Next, the function
$\dot{\theta}(\theta)$, obtained from \eqref{eq:ydot}-\eqref{eq:thetadot} for a
$y$-value corresponding to tumbling trajectories [Class-(A)], is plotted in
Fig. \ref{Fig2}(e). The range $\theta \in [3\pi,5\pi]$ used here, avoids
end-effects and corresponds to one spatial period seen in  Panel-1 of Fig.
\ref{Fig2}(b), where $y$ is evidently not constant.  Despite this, the
analytical plot captures the trend obtained from the simulations remarkably
well.  The plot indicates a relatively smaller average $\dot{\theta}$ when the
swimmer faces the wall ($\theta\in[4\pi,5\pi]$) than when it faces the
midsection ($\theta\in[3\pi,4\pi]$).  Consequently, the swimmer spends more
time oriented toward the wall and is thus driven on average in that direction
by its active velocity [{i.e.}, segment $ab < bc$ in Inset-1 of Fig.
\ref{Fig2}(b)].  The asymmetry in $\dot{\theta}$ is expected due to the
head-tail asymmetry about $\rb_s$ \cite{Hyd-cen} and the non-uniform shear rate
which increases from the midsection to the wall: when the swimmer faces the
wall, its `tail' is exposed to lower shear-rates, whereas the opposite is true
when the swimmer faces the midsection [see schematic, Fig. \ref{Fig2}(e)]. This
mechanism underlies the vertical migration described above.

The growing amplitude trajectories [Class-(B)] can be understood
similarly. A swimmer crossing $y=0$ encounters vorticity of opposite senses
across the midsection. For an HTS swimmer, the interplay of activity and
vorticity under this condition produces a spatial trajectory oscillating about
the midsection with constant amplitudes. Here, however, $\dot{\theta}$ has an
asymmetry that flips across the midsection, causing the swimmer to spend more
time facing the wall than the midsection in each channel half. This leads to
increasing amplitudes over time, implying $ab<cd$ in Inset-2, \ref{Fig2}(b).
Moreover, as amplitude grows, so does $\theta_m$, and when it reaches $\pi$,
the swimmer fully tumbles within the channel half, marking the onset of the
\emph{escaping and tumbling} phase of Class-B trajectories.

\begin{figure}

\includegraphics[width=\columnwidth]{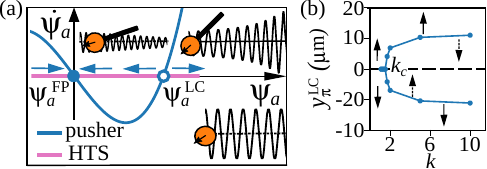}

\caption{(a) Schematic of $\dot\psi_a$ vs. $\psi_a$ showing stable (filled
circle) and unstable (open circle) fixed points for a \emph{pusher} and an HTS
microswimmer.  (b) Variation of the \emph{limit cycle} height, $y^{\rm{LC}}_\pi =
\left. y(\pi)\right|_{\rm{LC}}$ ($\approx y_{\rm{max}}$) [see Fig.
\ref{Fig2}(c)], as a function of flagellar flexibility.}

\label{Fig3}
\end{figure}

\textit{Limit cycle and trapping to the channel midsection -- }
Eliminating $y$ from Eq. \eqref{eq:thetadot} and expanding about the only
stable fixed point (FP), $(y^*, \theta^*) = (0, \pi)$, admitted by
\eqref{eq:ydot}-\eqref{eq:thetadot}, we obtain, up to the lowest nonlinear
order [Appendix A],
\begin{eqnarray}
\ddot{\psi} + \mu_1 \left(1 - \zeta\psi^2\right)\dot{\psi} + k_1 \psi = - \mu_3 \psi\dot{\psi}^2 - k_3 \psi^3.
\label{extended-VdP}
\end{eqnarray}
Here, $\psi$ and $\dot\psi$ denote \emph{small deviations} from the FP values
of $\theta$ and $\dot\theta$, respectively, with $\mu_1$, $k_1>0$ for a
\emph{pusher}. Notably, when the right-hand side equals zero, the equation
resembles the van der Pol equation (vdPE) differing only in the sign of
$\mu_1$. Despite these differences, for small $\mu_1/\sqrt{k_1} =\epsilon$,
$\mu_3/\zeta$, and $k_3/\zeta k_1$, Eq. \eqref{extended-VdP}, like vdPE, admits
the \emph{two-time} approximate solution $\psi(t^\prime) = \psi_a(\tau)
\cos{\left(t^\prime + \varphi(\tau)\right)}$, where $\tau = \epsilon t^\prime$
and $t^\prime = \sqrt{k_1}t$. Thus, the oscillations around the upstream
orientation ($\theta = \pi$) are sinusoidal on the fast time scale $t^\prime$,
while their amplitude, $\psi_a$, has a variation, evolving on a much slower
time scale $\tau$, akin to Class-B/C oscillations with growing/decaying
amplitude [Fig. \ref{Fig2}(b)].

Following \emph{Averaging Theory} \cite{strogatz2018nonlinear}, we obtain
$\dot{\psi}_a = -(\zeta\mu_1/8)\psi_a (4/\zeta - \psi_a^2)$. 
Thus, as shown in Fig.  \ref{Fig3}(a), $\psi_a$ has a \emph{stable} FP,
$\psi^{\rm{FP}}_a=0$, clearly corresponding to $(y^*, \theta^*)=(0,\pi)$. The
\emph{unstable} FP, $\psi^{\rm{LC}}_a$, represents a unique constant amplitude
oscillation ({\it i.e.}, a closed orbit in $y$-$\theta$ space), explaining the
\emph{unstable limit cycle} in Fig.  \ref{Fig2}(c). This further implies that
oscillations with $\psi_a<\psi^{\rm{LC}}_a$ exhibit decaying amplitude,
explaining Class-C trajectories [Panel-3, Fig.  \ref{Fig2}(b)], while those
with $\psi_a>\psi^{\rm{LC}}_a$ correspond to growing amplitude trajectories
(Class-B).

Remarkably, as shown in Appendix A, the solution $\psi_a(t)=2\{\zeta -
\exp{(\mu_t t)}[\zeta - 4 \psi^{-2}_a(0)]\}^{-1/2}$, obtained from above,
captures the qualitative rates of amplitude growth and decay seen in Fig.
\ref{Fig2}(b).  Additionally, we find that the phase $\varphi(t)$ varies with
time, unlike in vdPE,  providing a first-order explanation for the wavelength
variations observed along the trajectories in that figure.

\textit{Role of flexibility -- } For a flexible swimmer, the flagellar
bundle oscillates with nonzero $\phi$, increasing the average $\dot{\theta}$
and thus the frequency of $y$-$\theta$ trajectories as $k$ decreases. Since the
asymmetry in time spent `up' vs. `down' accumulates over cycles, more flexible
swimmers reach the walls faster for a given initial condition, as seen in Fig.
\ref{Fig2}(b). Furthermore, Fig. \ref{Fig3}(b) shows that as $k$ decreases the
unstable limit cycle enclosing decaying-amplitude trajectories undergoes a
\emph{subcritical} Hopf bifurcation, shrinking to an unstable FP at $k=k_c$,
consistent with the anomalous $k=2$ curve in Fig.  \ref{Fig2}(b),
\textit{Panel-3}.

In conclusion, we have demonstrated that head-tail shape-asymmetry
fundamentally drives active cross-streaming in channel flows. The asymmetry
induces a non-uniform rotation rate in the shear plane, which, combined with
self-propulsion, deflects microswimmers off-stream. Capturing this effect
requires extended hydrodynamic coupling to the imposed flow, though much of the
underlying physics can be understood via a simplified in-plane dynamics model
with minimally introduced asymmetric rotation rates. While HTA swimmers'
systematic approach to channel walls is noticeable but not fully explained in
previous studies \cite{junot2019swimming}, statistical effects can obscure
HTA-driven migration \cite{rusconi2014bacterial, Subramanian_shear_migration},
especially for short observation times [see Appendix C].

Our findings offer potential applications in sorting and controlling
microswimmer channel dynamics. For instance, center-trapped swimmers,
which are efficiently flushed out, can be separated from more flexible ones in a
mixed-flexibility population (see Fig. \ref{Fig3}(b), also \cite{flex_reason}).
Similarly, by tuning $v_f$, the maximum flow speed, swimmers can be separated
at the walls, with more rigid ones advancing further. Adjusting $v_f$ also
regulates the center-trapped fraction, increasing it up to $\sim
10\%$ of a rigid-swimmer population [Appendix B].

Our analysis extends naturally to HTA \emph{pullers}, which exhibit a
\emph{stable} limit cycle, unlike the \emph{unstable} one for \emph{pushers}.
The extended hydrodynamic coupling framework presented here is general and can
capture novel active-particle dynamics in any nonuniform flow, {\it e.g.},
those induced by microswimmers in dense suspensions.


\emph{Acknowledgments} -- TCA acknowledges grants CRG/2021/004759 from the
Science and Engineering Research Board (India) and MoE-STARS/STARS-2/2023-0814
for financial support. DCG acknowledges PMRF, Govt.  of India, for fellowship
and funding. TCA and DCG express gratitude to the \emph{Indian Institute of Science
Education and Research (IISER) Tirupati} for funds and facilities.
Furthermore, the support and the resources provided by `PARAM Brahma Facility'
under the National Supercomputing Mission, Government of India at the
\emph{Indian Institute of Science Education and Research (IISER) Pune} are
gratefully acknowledged.

\providecommand{\noopsort}[1]{}\providecommand{\singleletter}[1]{#1}%
%


\begin{thebibliography}{70}
\bibitem [{non()}]{nonextended}%
  \BibitemOpen
  \href@noop {} {}\bibinfo {note} {For rigid and symmetric particles, the
  coupling to external flows is through the flow vorticity and Fax\'{e}n's
  theorem, or its extended version \cite{kim_karilla}, which simply uses
  derivatives evaluated at the particle centers.}\BibitemShut {Stop}%
\bibitem [{she()}]{shear-induced}%
  \BibitemOpen
  \href@noop {} {}\bibinfo {note} {The cross-streaming in the shear-plane we
  discussed here is independent of, and can co-occur with, any shear migration
  due to chirality \cite{Stocker_bac_rheotaxis, Doi_chiral_migration}, if
  present.}\BibitemShut {Stop}%
\bibitem [{Hyd()}]{Hyd-cen}%
  \BibitemOpen
  \href@noop {} {}\bibinfo {note} {To be precise, the HT asymmetry about the
  hydrodynamic center $\rb_s - \rb_{\rm{sh}}$; however, the difference can be
  ignored for the present discussion due to the \emph{slenderness} of the
  flagellar rod.}\BibitemShut {Stop}%
\end{thebibliography}

\appendix
\renewcommand{\theequation}{A\arabic{equation}}
\setcounter{equation}{0}

\renewcommand{\thefigure}{A\arabic{figure}}
\setcounter{figure}{0}

\onecolumngrid


\vspace{2cm}

\twocolumngrid

\section{Appendix A: Solution of $\psi$ dynamics} 

Eq. \eqref{extended-VdP} involves the deviations $\psi = \theta - \theta^*$ and
$\dot \psi = \dot\theta
- \dot\theta^*$, where $\dot\theta^*$ denotes the value of $\dot\theta$ at the
  fixed point (FP). Rescaling $\psi\to x/\sqrt{\zeta}$, $t\to
t^\prime/\sqrt{k_1}$, and nondimensionalizing, we obtain, 
\begin{eqnarray}
x^{\prime\prime}   + x + \epsilon \left[ \left(1 - x^2\right)x^{\prime} 
                   + \frac{\epsilon_1}{ \epsilon}x x^{\prime 2} 
                   + \frac{\epsilon_2}{ \epsilon} x^3 \right] = 0.
\label{eq:vdp_x}
\end{eqnarray}
Here, $\epsilon_1 = \mu_3/\zeta$ and $\epsilon_2 = k_3/\zeta k_1$ are
dimensionless parameters, and the prime denotes differentiation with respect to
$t'$. For $\epsilon \ll 1$, with $\epsilon_1, \epsilon_2 \sim
\mathcal{O}(\epsilon)$, Eq.~\eqref{eq:vdp_x} describes a \emph{weakly nonlinear
oscillator}. It thus admits a perturbative solution of the form $x = x_0 +
\epsilon x_1 + \mathcal{O}(\epsilon^2)$, valid when at least two well-separated
timescales exist, {\it i.e.}, $x_0 = x_0(t', \tau)$ and $x_1 = x_1(t', \tau)$
with $\tau = \epsilon t'$ \cite{strogatz2018nonlinear}. Substituting $x =
x_0(t', \tau) + \epsilon x_1(t', \tau)$ into \eqref{eq:vdp_x}, we obtain
$x_0 = x_a(\tau) \cos(t' + \varphi(\tau))$, while requiring $x_1$ to remain
bounded at all times yields:
\begin{eqnarray}
\partial_\tau x_a\left( \tau \right) &=& \frac{x_a^3}{8} - \frac{x_a}{2}, \label{eq:xa}\\
\partial_\tau \varphi \left(\tau \right) &=& \frac{\sqrt{k_1}}{8\zeta\mu_1} \left(\mu_3 
                              + 3\frac{k_3}{k_1}\right)x_a^2. \label{eq:phitau}
\end{eqnarray}    
Solving the above gives $x_a(\tau)$ and $\varphi(\tau)$, which, in terms of
the original variables, read as:
 \begin{eqnarray}
&&\psi_a\left(t\right) = \frac{2/\sqrt{\zeta}}{ \sqrt{1 + e^{\mu_1 t}\left[\left(\frac{2/\sqrt{\zeta}}{\psi_a\left(0\right)}\right)^2 -1\right]}  }, \label{eq:psia}\\
&&\varphi\left(t\right) = \varphi\left(0\right) + \frac{\mu_3 + 3\left(k_3/k_1\right)}{2 \zeta } \frac{\sqrt{k_1}}{\mu_1} \times  \nonumber \\
&&\left[\mu_1 t -  \ln{ \left| \left(\frac{\zeta\psi^2_a\left(0\right)}{4} -1 \right)e^{\mu_1 t} - \frac{\zeta\psi^2_a\left(0\right)}{4} \right|}\right]. \label{eq:varphi}
\end{eqnarray}
Thus, at zeroth order, $x$, or equivalently $\psi$, exhibits sinusoidal
oscillations on the timescale $t^\prime$ with amplitude and phase slowly
varying on the much longer timescale $\tau$. Eqs. \eqref{eq:psia} and
\eqref{eq:xa}, expressed in the original variables, give the forms of $\psi_a$ and $\partial_t \psi_a \equiv \dot{\psi}_a$, respectively, appearing in the main text.

\begin{figure}

\includegraphics[width=0.8\columnwidth]{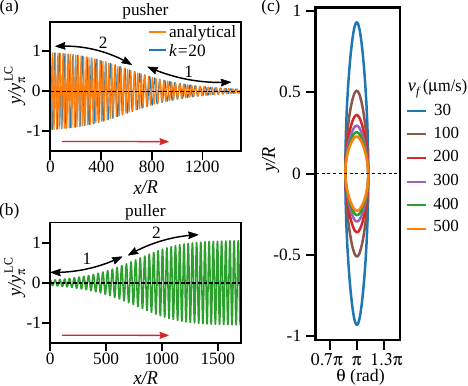}

\caption{(a) Scaled trajectories for rigid swimmers from the analytical
solution (orange) and full simulations for $k=20$ (overlapping blue), starting
at $(y_i/y^{\rm{LC}}_\pi, \theta_i) = (0.97, \pi)$.  (b) Simulation trajectory
of a puller with $k=20$, normalized by the corresponding $y^{\rm{LC}}_\pi$ for
a pusher. Labels 1 and 2 indicate regions of differing curvature in both (a)
and (b). Red arrows show flow/swimming direction.  (c) LC for rigid swimmers
($k=20$) from full simulations for varying flow speeds $v_f$, with $v_s = 50\,
\mu\mathrm{m}/\mathrm{s}$.}

\label{Fig1_Appendix}

\end{figure}

\emph{Trajectory solutions --} From the above, we have
\begin{eqnarray}
\theta\left(t\right) = \pi + \psi \left(t\right) \approx \pi + \psi_a\left(t\right) \cos\left(\sqrt{k_1} t + \varphi\left(t\right)\right),
\label{eq:theta_sol}
\end{eqnarray}
which also relates $\{\psi_a(0), \varphi(0)\}$, appearing in \eqref{eq:psia}
and \eqref{eq:varphi}, to the initial conditions $\{\dot\theta_i, \theta_i\}$
or $\{y_i, \theta_i\}$. Substituting $\theta(t)$ and the corresponding
$\dot\theta(t)$ into Eq.~\eqref{eq:thetadot} yields $y(t)$. For large $v_f$,
approximating $x \approx v_f t$, we substitute $t = x/v_f$ to obtain $y(x)$,
which is plotted in Fig.~\ref{Fig1_Appendix}(a). The plot compares $y(x)/y^{\rm
LC}_\pi$ from the simplified theory with that obtained from full simulations.
The normalization factor $y^{\rm LC}_\pi = \left. y(\pi) \right|_{\rm LC}$
accounts for variations in $y_{\rm LC}$ with and without self-induced active
flows.

For pullers, the $\psi$-equation becomes $\ddot{\psi} - \mu_1(1 -
\zeta\psi^2)\dot{\psi} + k_1 \psi = -\mu_3 \psi \dot{\psi}^2 - k_3 \psi^3$,
where the sign reversal of the $\dot{\psi}$ term transforms the limit cycle
into a stable one. This leads to growing amplitude trajectories inside the
limit cycle, in contrast to the decaying ones seen for pushers, as shown in
Fig.~\ref{Fig1_Appendix}(b). In fact, for our model, a puller's motion along
$y$ and $\theta$ axes maps to that of a pusher moving backwards in time, while
along the $x$ axis they both move in the same direction.

\emph{Limit cycle (LC) -- } From Eq.~\eqref{eq:psia}, $\psi_a(0) =
2/\sqrt{\zeta} \equiv \psi_a^{\rm LC}$ is the constant amplitude on the LC,
with the corresponding phase $\varphi(t)$ obtained from Eq.~\eqref{eq:varphi}.
Thus, $\theta(t)$ on the LC is given by
\begin{eqnarray}
\theta^{\rm LC} = \pi + \psi_a^{\rm LC} \cos\left(\omega^{\rm LC} t + \varphi^{\rm LC}\right),
\label{eq:theta_LC}
\end{eqnarray}
where the rescaled frequency is $\omega^{\rm LC} = \sqrt{k_1}\left[1 + (\mu_3 +
3k_3/k_1)/2\zeta\right]$ and the phase constant is $\varphi^{\rm LC} =
\cos^{-1}\left[(\theta_i - \pi)/\psi_a^{\rm LC}\right]$. For completeness,
$y^{\rm LC}$ can be obtained by the same procedure used to derive $y(t)$ from
$\theta(t)$, yielding $y^{\rm LC}(\theta^{\rm LC})$. However, we can estimate
its approximate shape as follows. Assuming a narrow $\theta$-width of the LC,
Eq.~\eqref{eq:thetadot} gives $y^{\rm LC} \approx R^2 \dot\theta^{\rm LC}/v_f
\alpha$. 
Hence, using \eqref{eq:theta_LC} we easily see
that the LC is nearly an \emph{ellipse} surrounding the FP $(0,\pi)$ given by
\begin{eqnarray}
\frac{\left(y^{\rm LC}\right)^2}{a^2} + \frac{\left(\theta^{\rm LC} - \pi\right)^2}{b^2}
= 1, \nonumber
\end{eqnarray} 
with $a = \psi_a^{\rm LC} \omega^{\rm LC} R^2/v_f \alpha $ and $b =
\psi_a^{\rm LC}$.

\emph{Motion around LC --} The evolution of $\theta_a(t)$, the amplitude of $\theta(t)$, starting from
small deviations from LC, follows from \eqref{eq:psia} as: $\theta_a(t) \approx
\pi + \psi_a^{\rm{LC}}\left(1 - g e^{\mu_1 t}/2\right)$ for small
$t$, and $\theta_a(t) \approx \pi + \left(\psi_a^{\rm{LC}}/\sqrt{g}\right)
e^{-{\mu_1}t/2}$ for large $t$, where $g =
\left[\psi_a^{\rm{LC}}/\psi_a(0)\right]^2 - 1$. These describe the curvature
change of decay profiles within the LC ($g > 0$) for short and long times
(regions 1 and 2 in Fig.~\ref{Fig1_Appendix}(a),(b)), and growth outside the LC
($g < 0$) for short times.

\section{Appendix B: Swimmer population flushed through channel midsection}  
The LC separates trajectories trapped to the midsection from those
escaping to the walls.  Therefore, assuming an initial population uniformly
distributed in $y$-$\theta$ space, and neglecting interactions and noise, the
fraction of population flushed through the channel midsection, is given by the
ratio of the area $\mathcal{A}$ of LC  to the total phase space area $2\pi\times 2R$.
As shown in Fig. \ref{Fig1_Appendix}(c), $\mathcal{A}$ can be tuned by
adjusting the imposed flow speed $v_f$, reaching a maximum height
$y^{\rm{LC}}_\pi = R$.
With the approximation of the LC as an ellipse (see above), for any given
$v_f$, we estimate $\mathcal{A} = \pi a b$ with $a$ and $b$ measured from
\ref{Fig1_Appendix}(c). Thus, for low enough $v_f$, the maximum trapped
fraction is estimated as $\mathcal{A}_{\rm{max}}/4\pi R \approx \pi R~(0.13\pi)/4\pi R
\approx 0.10$, where $\mathcal{A}_{\rm{max}} = \pi a_{\rm{max}} b_{\rm{max}}$, with
$a_{\rm{max}} = R$ and $b_{\rm{max}}=0.13\pi$ Thus, by tuning $v_f$, up to 10\%
of the population can be flushed.

\begin{figure}[t]

\includegraphics[width=\columnwidth]{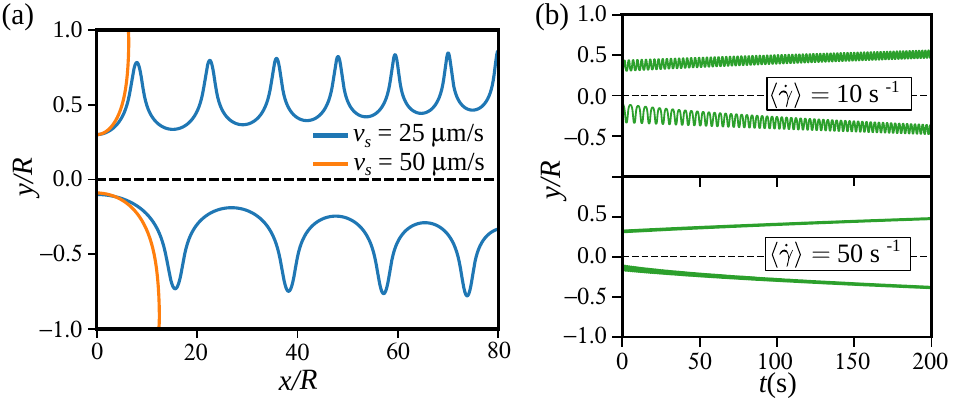}

\caption{(a) Spatial trajectories from two representative initial heights,
$y_i/R = 0.3$ and $-0.1$, using parameters similar to those in
Ref.~\cite{junot2019swimming}.  (b) $y/R$ vs. $t$ for parameters similar to those in
Ref.~\cite{rusconi2014bacterial}.  In both (a) and (b), $k = 20$ (rigid
swimmers) and $\theta_i = 0$ are chosen to yield tumbling trajectories.}

\label{Fig2_Appen}

\end{figure}

\section{Appendix C: Comparison with previous studies}
Figure~\ref{Fig2_Appen}(a) shows trajectories of a rigid swimmer using
parameters similar to those in \emph{Junot et al.}~\cite{junot2019swimming}: $R =
50\,\mu\mathrm{m}$, $v_f = 200\,\mu\mathrm{m}/\mathrm{s}$, and hence  an average
shear rate $\langle \dot\gamma \rangle = v_f / R = 4\,\mathrm{s}^{-1}$. As
predicted in the main text, the trajectories exhibit clear slopes toward the
walls. Remarkably, they closely correspond to some of the
experimental trajectories reported in the reference.

Similarly, Fig.~\ref{Fig2_Appen}(b) shows rigid swimmer trajectories over
time using parameters similar to those in \emph{Rusconi et al.}
\cite{rusconi2014bacterial}: $R = 200\,\mu\mathrm{m}$ and two $v_f$ values
corresponding to $\langle \dot\gamma \rangle = 10\,\mathrm{s}^{-1}$ and
$50\,\mathrm{s}^{-1}$. The plot indicates that HTA-induced shear migration is
not significant within the $\sim 60\,\mathrm{s}$ typical observation window
reported in the study. Thus, the swimmer migration without HTA effects,
presented in that work, is consistent with our analysis.

Finally, we note that the typical time to reach the wall $t_{\rm{wall}}\sim
R/\langle \dot y \rangle_t$, with $\langle\cdots\rangle_t$ representing time
average over one period, scales as  $R^2$ divided  by a factor linear in $v_s$
and $\langle \dot\gamma\rangle$.  Therefore, given similar values of $v_s$ and
$\langle \dot\gamma\rangle$, a change in $R$ significantly alters
$t_{\rm{wall}}$. Specifically, for $R =50 \, \mu \rm{m}$, $\langle
\dot\gamma\rangle= 4 \,\rm{s}^{-1}$ and $v_s=25\, \mu\rm{m}/\rm{s}$, used in
\cite{junot2019swimming}, we estimate $t_{\rm{wall}}\approx 100\, \rm{s}$,
while for $R =200 \, \mu \rm{m}$, $\langle \dot\gamma\rangle= 10
\,\rm{s}^{-1}$ and $v_s=25\, \mu\rm{m}/\rm{s}$, used in
\cite{rusconi2014bacterial}, we estimate $t_{\rm{wall}}\approx 1400\, \rm{s}$.

\end{document}